\documentclass{desyproc}

\begin{document}
\title{pp scattering in t and b spaces}

\author{{\slshape A. K. Kohara, E. Ferreira, T. Kodama  }\\[1ex]
  Instituto de Fisica, Universidade Federal do Rio de Janeiro, Rio de Janeiro, Brazil \\
 }

\contribID{smith\_joe}



\maketitle

\begin{abstract}
We propose analytical forms, in both momentum transfer and impact parameter    
spaces,  for the amplitudes of elastic pp scattering,  giving coherent and     
accurate description of the observables  at all energies  $\sqrt{s}\geq 20$ GeV.   
The real and imaginary parts are separately identified through their roles            
at small and large t values.  The study of the differential cross sections in b-space    
leads to  the understanding of the effective interaction ranges contributing to     
elastic and inelastic processes.                                                        
\end{abstract}

\section{Amplitudes   }
\begin{figure*}[hb]
\centerline{\includegraphics[width=0.45\textwidth]{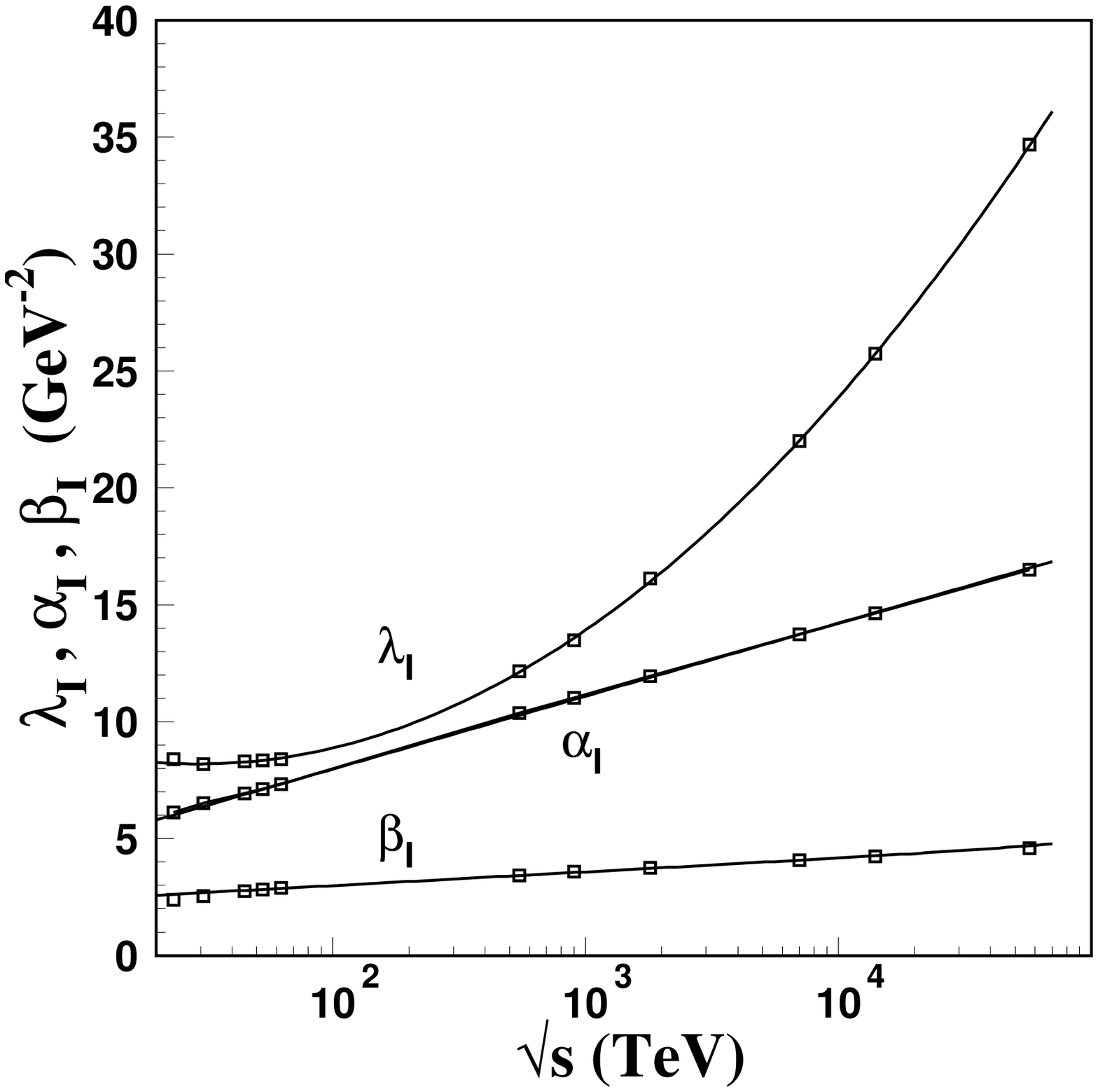}}
\caption{The sum of the parameters $\alpha_I$ and $\lambda_I$ determine 
the pp total cross section through the optical theorem. Their energy 
dependences (linear and quadratic respectively) are determined with 
accuracy in the whole energy interval. The parameter $\beta_I$ is also shown.}
\label{FigureOne}
\end{figure*}

The proposed amplitudes in t and b spaces, connected by Fourier transforms, are 
written 
\begin{equation}
T_{K}^{N}(s,t)=\alpha_{K}(s)\mathrm{e}^{-\beta_{K}(s)|t|}+\lambda_{K}%
(s)\Psi_{K}(\gamma_{K}(s),t)+\delta_{K,R}R_{ggg}\left(  t\right)
~,\label{ampTK}%
\end{equation}
with the t-shape function 
\begin{equation}
\Psi_{K}(\gamma_{K}(s),t)=2~\mathrm{e}^{\gamma_{K}}~\bigg[{\frac
{\mathrm{e}^{-\gamma_{K}\sqrt{1+a_{0}|t|}}}{\sqrt{1+a_{0}|t|}}}-\mathrm{e}%
^{\gamma_{K}}~{\frac{e^{-\gamma_{K}\sqrt{4+a_{0}|t|}}}{\sqrt{4+a_{0}|t|}}%
}\bigg]~,\label{cd1_s}%
\end{equation}
and 
\begin{equation}
\tilde{T}_{K}(s,b)=\frac{\alpha_{K}}{2~\beta_{K}}e^{-\frac{b^{2}}{4\beta_{K}}%
}+\lambda_{K}~\tilde{\psi}_{K}(s,b)~,
\end{equation}
with the b-shape function
\begin{equation}
\tilde{\psi}_{K}(s,b)=\frac{2~e^{\gamma_{K}}}{a_{0}}~\frac{e^{-\sqrt
{\gamma_{K}^{2}+\frac{b^{2}}{a_{0}}}}}{\sqrt{\gamma_{K}^{2}+\frac{b^{2}}%
{a_{0}}}}\Big[1-e^{\gamma_{K}}~e^{-\sqrt{\gamma_{K}^{2}+\frac{b^{2}}{a_{0}}}%
}\Big]~. 
\end{equation}
The parameters (four for the imaginary and four for the real parts) have very 
regular behavior. In Figure \ref{FigureOne}  we show the parameters $\alpha_I$ (linear in 
$\log{s}$) and $\lambda_I$ (quadratic in $\log{s}$ 
of the imaginary part, that together determine the total cross section. 

The normalization is defined by 
\begin{equation}
\frac{d\sigma}{dt}= \left(  \hbar c\right)  ^{2}~|T_{R}(s,t)+iT_{I}%
(s,t)|^{2}~~,~~~ \sigma= 4 \sqrt{\pi} \left(  \hbar c\right) ^{2}~ T_{I}^{N}(s,t=0) ~.
\end{equation}
 These amplitudes account for the non-perturbative dynamics governing   elastic 
scattering  at high energies   \cite{EF1,KEK1,KEK2} .
The imaginary and real amplitudes in t-space show respectively one and two zeros.
For very large t above 5 GeV$^2$ there are additional  perturbative contributions
to the real part, of very small magnitude (three gluon exchange $R_{ggg}$). 

  \section{Interaction Ranges}
The differential cross sections $d\hat \sigma_{\rm elastic}/d^2\vec{b}$, etc, 
that correspond to interactions in rings of radius b in the transverse plane 
are written 
\begin{eqnarray}
\label{Sigma_elas}
\sigma_{\rm elas}=\frac{(\hbar c)^2}{\pi}\int d^2\vec b~ |\tilde T(\vec b,s)|^2 =\int d^2\vec b ~\frac{d\hat \sigma_{\rm elas}}{d^2\vec b} ~ , 
\end{eqnarray}
\begin{eqnarray}
\label{Sigma_total_H_Ar}
\sigma_{\rm tot}&=&4\sqrt\pi(\hbar c)^2~T_I(s,t=0)=\frac{2}{\sqrt\pi}(\hbar c)^2\int d^2\vec b ~\tilde T_I(s,\vec b)=\int d^2\vec b ~\frac{d\hat \sigma_{\rm tot}}{d^2\vec b}~, 
\end{eqnarray} ~   
\begin{eqnarray}
\sigma_{\rm inel}=(\hbar c)^2\int d^2\vec b ~
\big( \frac{2}{\sqrt\pi}\tilde T_I(s,\vec b)-\frac{1}{\pi}|\tilde T(\vec b,s)|^2\big) ~  
=\int d^2\vec b~ \frac{d\hat \sigma_{\rm inel}}{d^2\vec b}~. 
\end{eqnarray}

\begin{figure*}[hb]
{\includegraphics[width=0.48\textwidth]{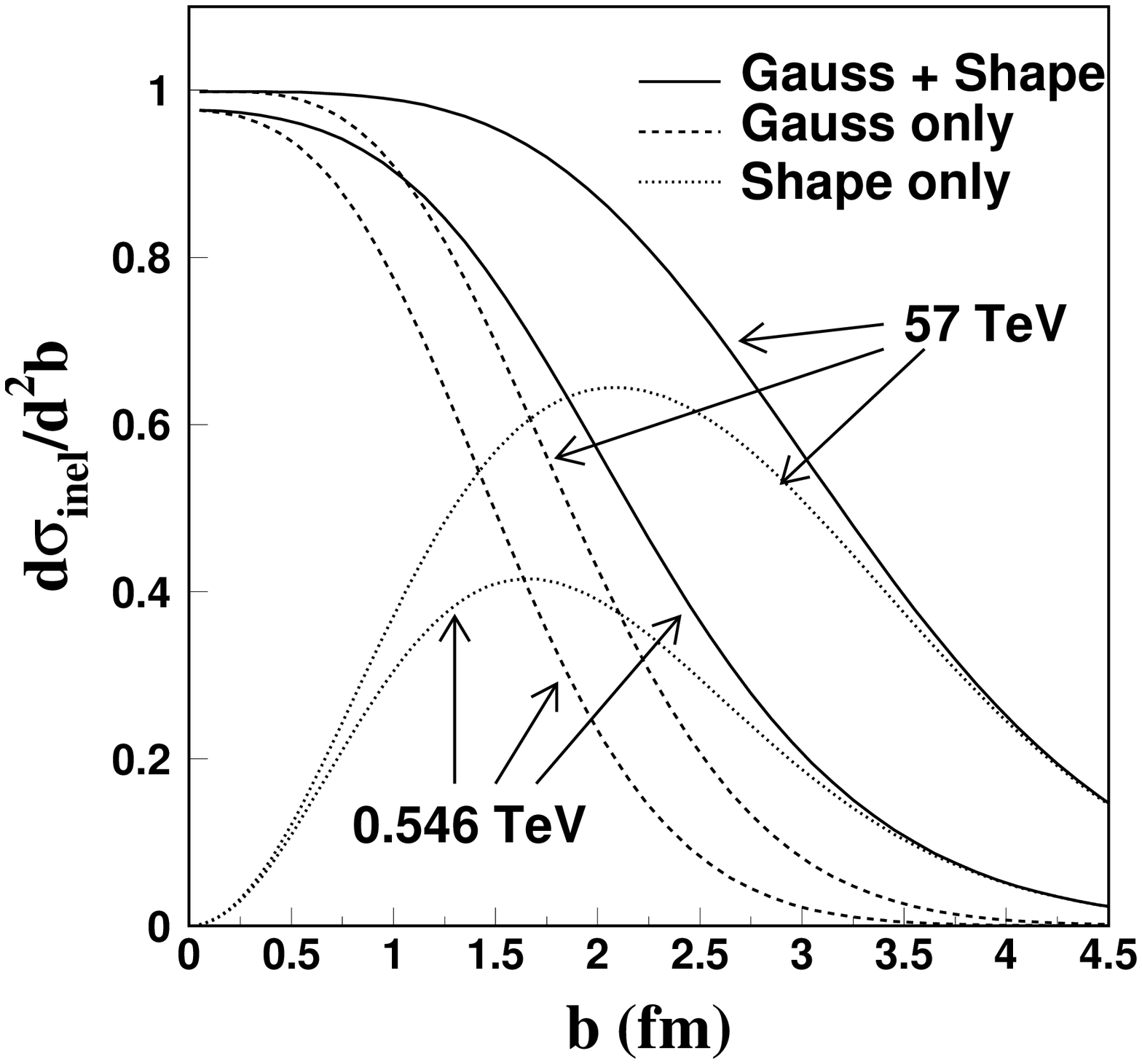}}
 {\includegraphics[width=0.46\textwidth]{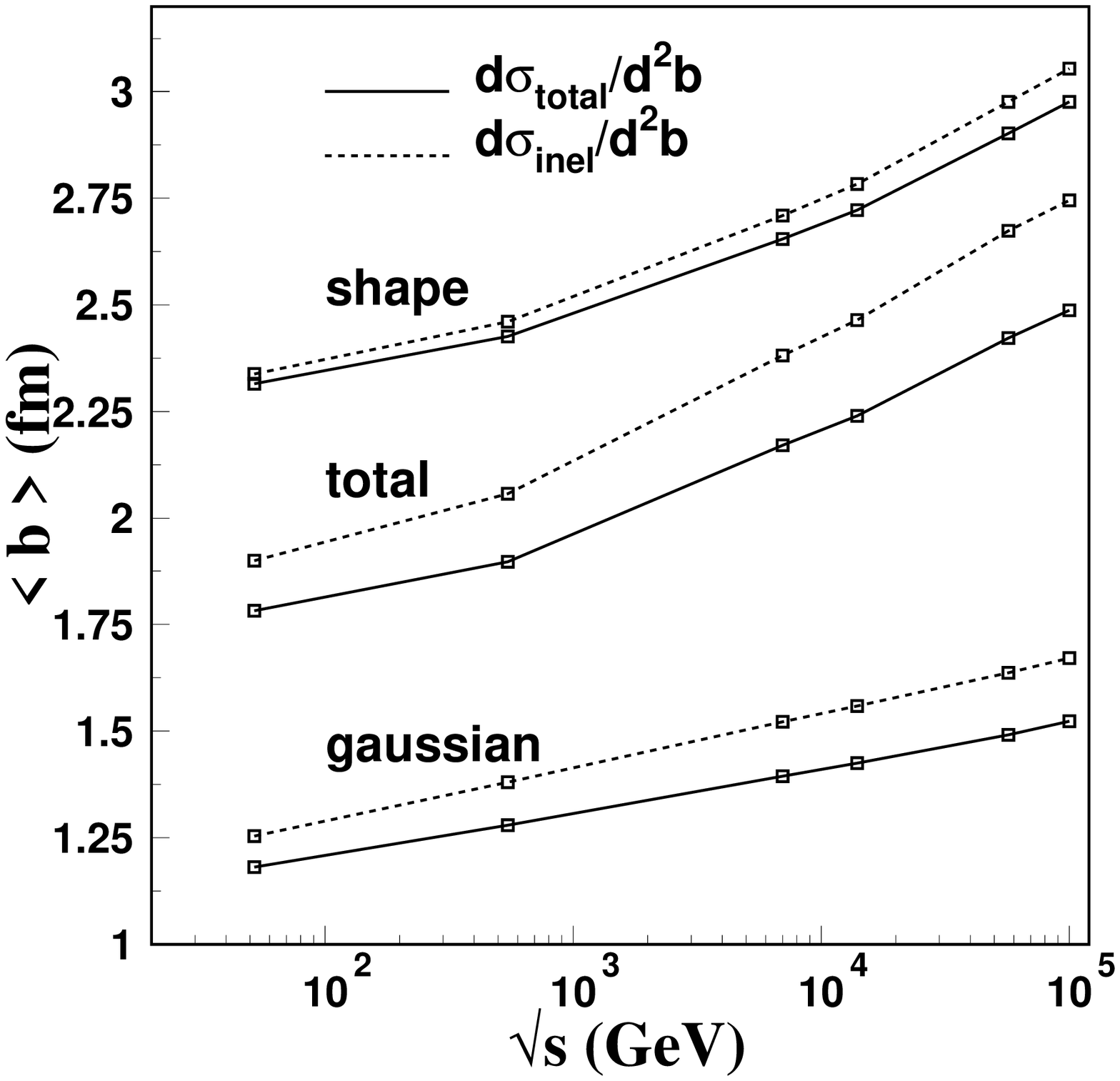}}
\caption{1) b-dependence of the contributions of the Gaussian, shape 
and total amplitudes to the b-differential inelastic cross section.
The plot compares the behaviour at 546 GeV and 7 TeV ;   
2) Average values of the interaction ranges related to the  
Gaussian  and Shape parts, each taken separately,  and 
related to the whole amplitudes with both parts. We observe that the 
Shape function leads to interactions that have  longer ranges 
and stronger energy increases (like quadratic in $\log{s}$). The 
Gaussian interaction ranges increase only linearly with $\log{s} . $
}
\label{FigureTwo}
\end{figure*}

In the LHS of Figure \ref{FigureTwo}  we show the  forms of the   inelastic differential  
cross sections as function of b  for the energies 546 GeV and 7 TeV. These functions are 
also shown separately for the cases of pure contributions from the Gaussian part and from 
the Shape function part of the amplitude. 
We observe that the Gaussian part is more concentrated for small b, with limited increase with 
the energy, and that the shape function part is extended for larger b values, and 
   that  it  determines the observed energy increase of the integrated cross section 
(squared $\log{s}$). 
 The distinguished behaviour of the two parts of the amplitude (Gaussian and Shape Function parts)
is quantitatively shown in the RHS of Figure \ref{FigureTwo}, where the average values of the 
interaction radius are represented. These quantities are 
\begin{eqnarray}
\langle b \rangle = \frac{1}{N}\int b\frac{1}{(\hbar c)^2}\frac{d\sigma_{inel}}{d^2\vec b} d^2\vec b = \frac{1}{N}\int b~G(s,b) d^2\vec b~ .
\end{eqnarray}
with the normalization factor N. Analogous expressions are defined 
for the total and elastic cross sections.
\section{Acknowledgments}
We acknowledge support received from the Brazilian governamental 
agencies  CNPq and FAPERJ. 
 

\begin{footnotesize}

\end{footnotesize}
\end{document}